\documentclass[runningheads,a4paper]{llncs}

\PassOptionsToPackage{hyphens}{url}
\usepackage{hyperref}
\usepackage{graphicx}
\usepackage{calc}
\usepackage{amssymb}
\usepackage{amstext}
\usepackage{amsmath}
\usepackage{url}
\usepackage{placeins}
\usepackage{multirow}
\usepackage{caption}
\usepackage{subcaption}
\usepackage{hhline}

\usepackage{listings}
\begin{document}

\title{An Extended Evaluation Split for DeepSpaceYoloDataset}
\titlerunning{An Extended Evaluation Split for DeepSpaceYoloDataset}

\author{Olivier Parisot}
\authorrunning{O. Parisot}
\institute{
	Luxembourg Institute of Science and Technology (LIST) \\ 
	5 Avenue des Hauts-Fourneaux, \\
	4362 Esch-sur-Alzette, Luxembourg \\ 
	\email{olivier.parisot@list.lu}
}


\maketitle              

\begin{abstract}

Recent technological advances in astronomy, particularly the growing popularity of smart telescopes for the general public, make it possible to develop highly effective detection solutions that are accessible to a wide audience, rather than being reserved for major scientific observatories.
Published in 2023, DeepSpaceYoloDataset is a collection of annotated images created to train YOLO-based models for detecting Deep Sky Objects, particularly suited for Electronically Assisted Astronomy.
In this paper, we present an update to DeepSpaceYoloDataset with the addition of a new split, test2026, designed to evaluate detection models with a greater diversity of images.

\end{abstract}

\section{Introduction}

Astronomical imaging is expanding rapidly, driven both by major surveys conducted at large professional observatories, such as those currently in operation and those planned for the near future \cite{reitze2024evolution}, and by the growing activity of astronomy enthusiasts \cite{sule2025astronomy}, enabled by equipment that is becoming more affordable and easier to use.
The development of autonomous telescopes is now a reality \cite{mehta2025autonomous}, and the growing availability of tiny smart telescopes further enhances the relevance of underlying technologies for the amateur astronomy community as well as for a wider audience.

In recent years, work has been carried out to propose Deep Learning techniques for performing robust detection in astronomical images, in order to automatically identify targets of interest, across diverse conditions, instruments, and noise levels. 
Introduced in 2023, DeepSpaceYoloDataset is a dataset designed to train supervised Deep Learning models for the detection of DSO (Deep Sky Objects: galaxies, nebulae, and star clusters), using architectures such as YOLO or RT-DETR (lightweight versions, such as \textit{nano} models, are particularly suitable for Electronically Assisted Astronomy).
It is a set of 4,696 RGB images collected between March 2022 and September 2023 with two smart telescopes (Stellina and Vespera) and annotated with the positions of visible DSO.
This dataset was widely used by scientific studies such as \cite{kanev2025contrast}, and enriched with data augmentation by \cite{ramos2025deep}.
A similar dataset has even been published recently \cite{piratinskii2025cosmica}.

For these studies, standard dataset splits (train/validation/test) were commonly used to evaluate model performance.
In \cite{parisot2025robustnessanalysisdeepsky}, we had notably proposed splitting the images according to their filenames, resulting in a distribution of 4,252 images for training, 333 for validation, and 222 for testing.
In fact, our experiments suggest that the original test split may introduce evaluation inconsistencies, leading to underestimated and less stable performance measurements.
Among the current limitations, we can highlight the following:
\begin{itemize}
\item The images included in the test split are no longer representative of the significant improvements achieved by smart telescope manufacturers, particularly regarding the reduction of walking noise, as well as the ability to automatically stack images of the same target across multiple nights, which greatly enhances image quality.
\item The test set does not include galaxy images and contains very few globular clusters; as a result, models are primarily evaluated based on their ability to detect nebulous structures.
\end{itemize}

In this work, we introduce an additional evaluation split, \textit{test2026}, designed to provide a more representative and consistent benchmark.

\section{Dataset Update}

The original version of DeepSpaceYoloDataset consists of 4,696 images organized into standard YOLO format directories \cite{parisot2024deepspaceyolodataset}.

\begin{figure}[htbp]
    \centering
    \includegraphics[width=0.7\linewidth]{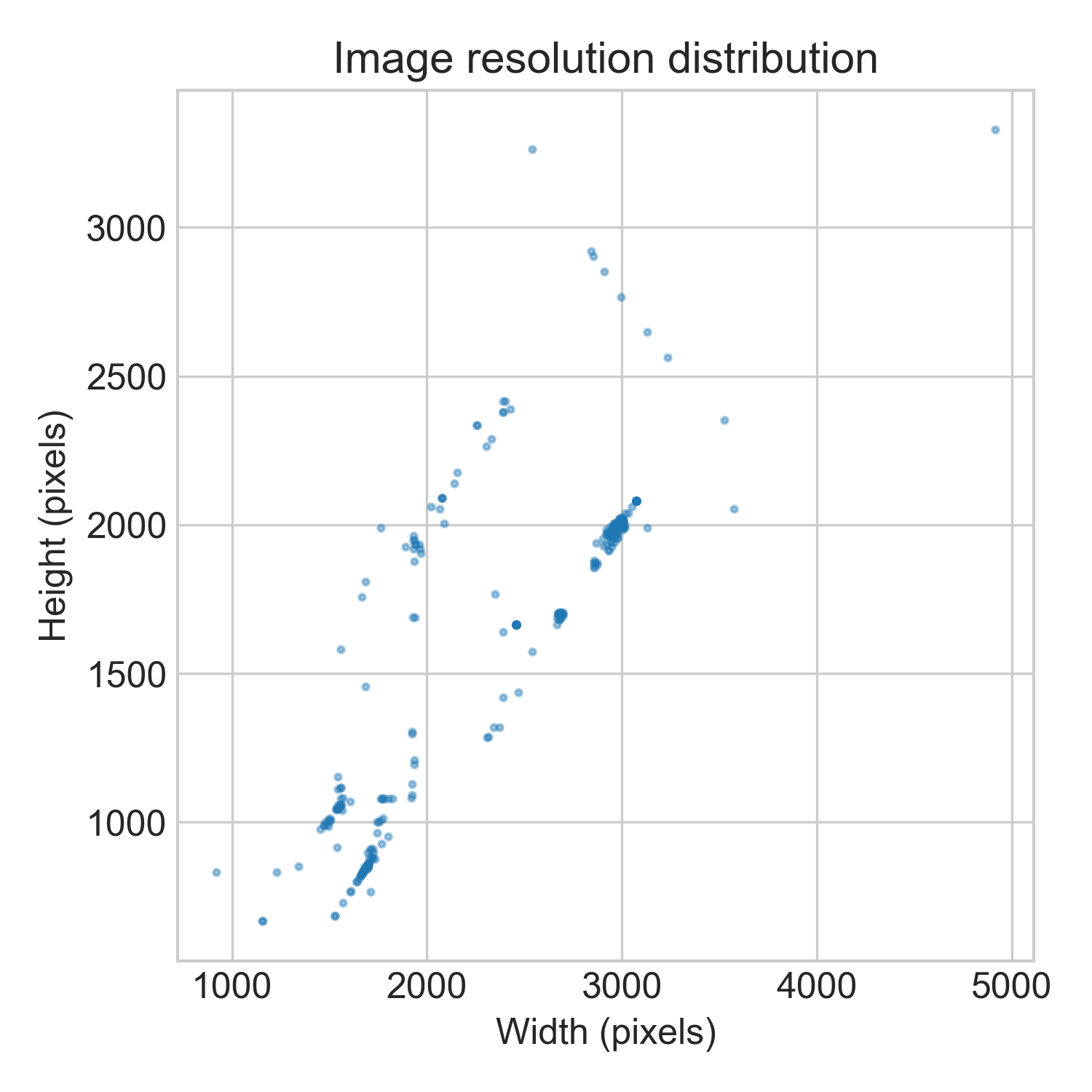}
    \caption{Distribution of image dimensions (width, height) in the \textit{test2026} split.}
    \label{fig:stats}
\end{figure}

We have extended this dataset by introducing a new split called \textit{test2026} with additional 335 images with different high-resolution (Figure \ref{fig:stats}).
In this split, the median image resolution is 2681x1927 pixels (mean resolution 2434x1661). 
The aspect ratio distribution is centered around 1.52 (o = 0.26). 
Most images are landscape-oriented (89.3\%), with 1.2\% portrait and 9.6\% approximately square images.

The dataset now contains 5,031 images, and here is its file organization structure:

\begin{verbatim}
DeepSpaceYoloDatasetV2/
  train/ 	(4252 images of 608x608 pixels)
  val/ 		(333 images of 608x608 pixels)
  test/ 		(222 images of 608x608 pixels)
  test2026/ 	(335 images of different resolutions)
\end{verbatim}

All annotations follow the YOLO format (one text file per images, with the specifications of bounding boxes) and remain fully compatible with existing training and evaluation pipelines.

In the following subsections, we explain how we have collected and processed the additional data.

\subsection{Data collection}

Following the same approach as for the first version of the DeepSpaceYoloDataset, images were captured until April 2026 using the same smart telescopes (Stellina and Vespera) in the same geographical region (France, Belgium, and Luxembourg).
The protocol described in \cite{parisot2023milan} was followed; its main steps are summarized below:
\begin{itemize}
\item Default parameters of Stellina and Vespera were applied: 10 seconds for exposure time per frame and 20 dB for gain.
\item The retained images correspond to the outputs directly produced by the telescopes after filtering, alignment, and stacking of the raw frames, acquired with moderate cumulative integration times (20–120 minutes), typically sufficient to achieve a good signal-to-noise ratio for most targets.
\end{itemize}

In practice, by aggregating images over the 2021–2026 period, we have collected images of at least 335 targets, including DSO visible from the Northern Hemisphere (galaxies, open and globular clusters, emission/reflection/planetary nebulae, etc.) \cite{steinicke2010observing}, but also comets like \textit{12P Pons-Brooks} and \textit{C2022 E3 ZTF}.
The selection of targets was based on a selection performed with the Stellarium software \cite{stellarium2026}. 
This preselection considered the portion of the night sky that was visible and free of clouds at the time of observation, and the magnitude of the targets was also taken into account.

Obtaining decent astronomical observations was ofter challenging due to frequent unfavourable weather conditions and the increasing proliferation of satellites (which produced unwanted streaks in many images). 
As a result, we manually inspected the images and applied a filtering step ourselves to discard those whose quality was too degraded.

\subsection{Processing}

\begin{figure}[htbp]
    \centering
    \includegraphics[width=0.95\linewidth]{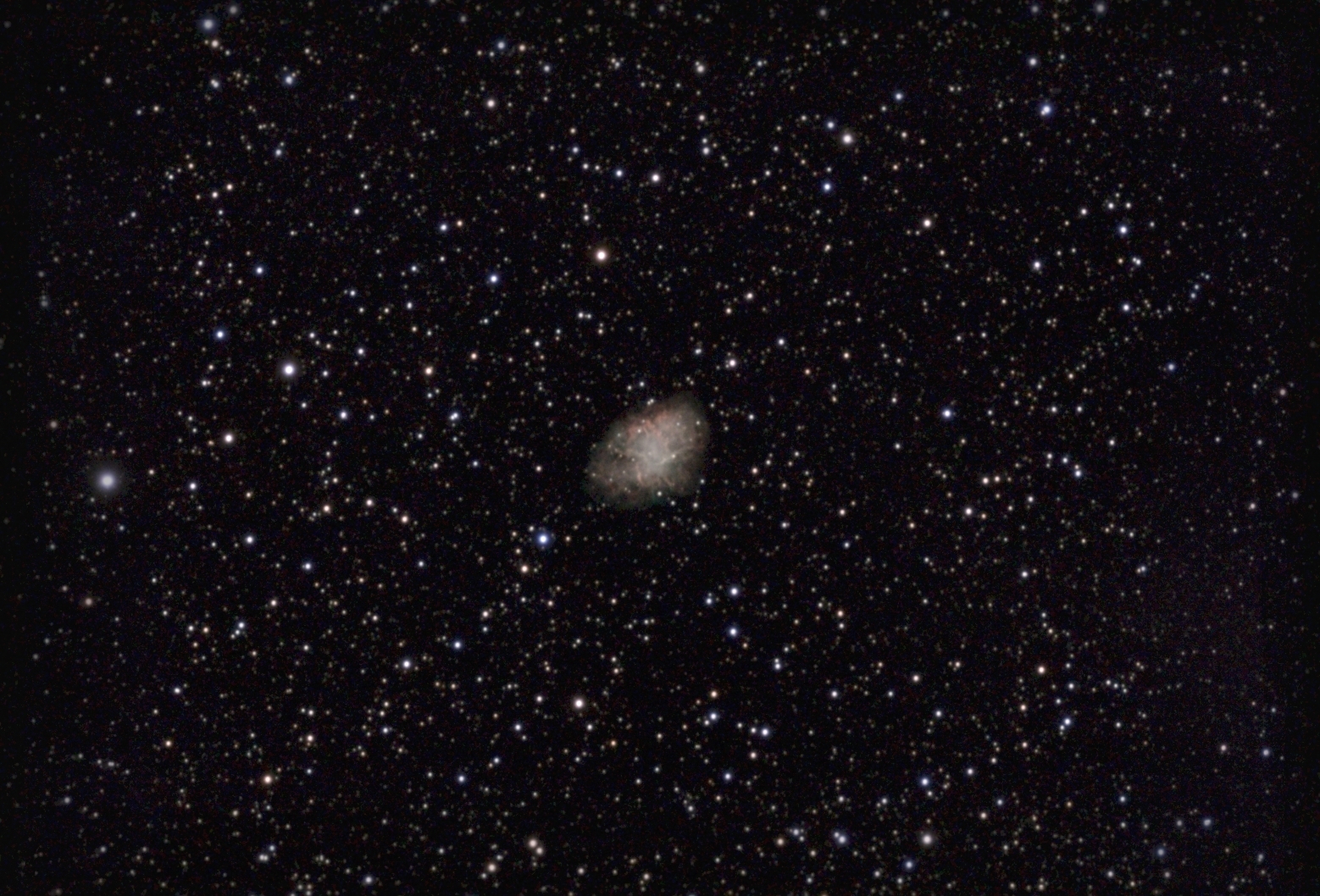}
    \caption{Composite image of M1 supernova remnant, captured with a Stellina and a Vespera: 1752 frames of 10 seconds captured during several nights (1/3/2022, 2/3/2022, 16/12/2022, 8/2/2023, 3/4/2023, 5/11/2023, 17/12/2023, 19/3/2024, 27/12/2024, 28/12/2024, 3/2/2025).}
    \label{fig:m1}
\end{figure}

\begin{figure}[htbp]
    \centering
    \includegraphics[width=0.95\linewidth]{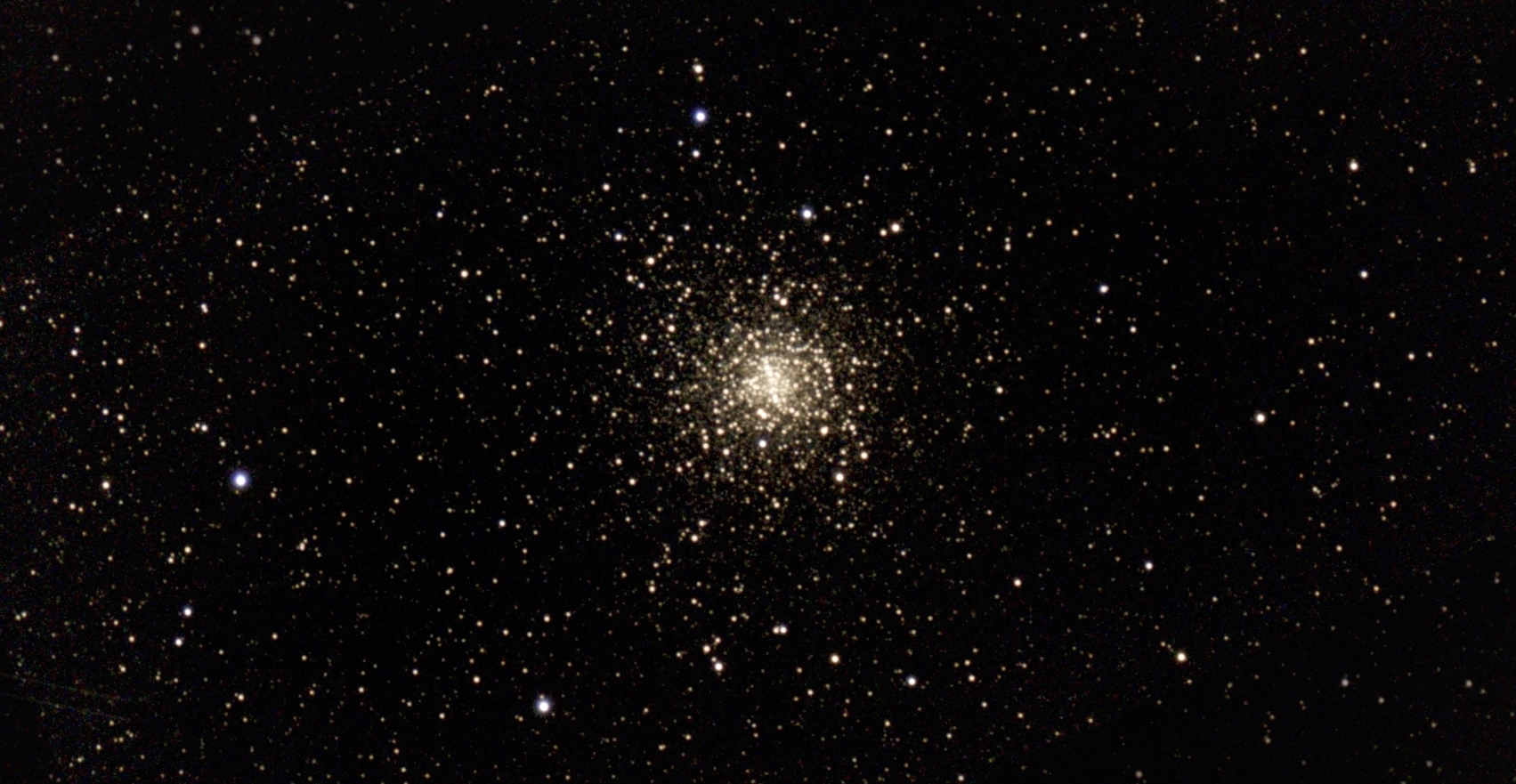}
    \caption{Composite image of M4 globular cluster, captured with a Stellina and a Vespera: 776 frames of 10 seconds captured during several nights (4/7/2023, 30/6/2024, 20/8/2024, 28/6/2025, 29/6/2025, 18/8/2025).}
    \label{fig:m4}
\end{figure}

\begin{figure}[htbp]
    \centering
    \includegraphics[width=0.95\linewidth]{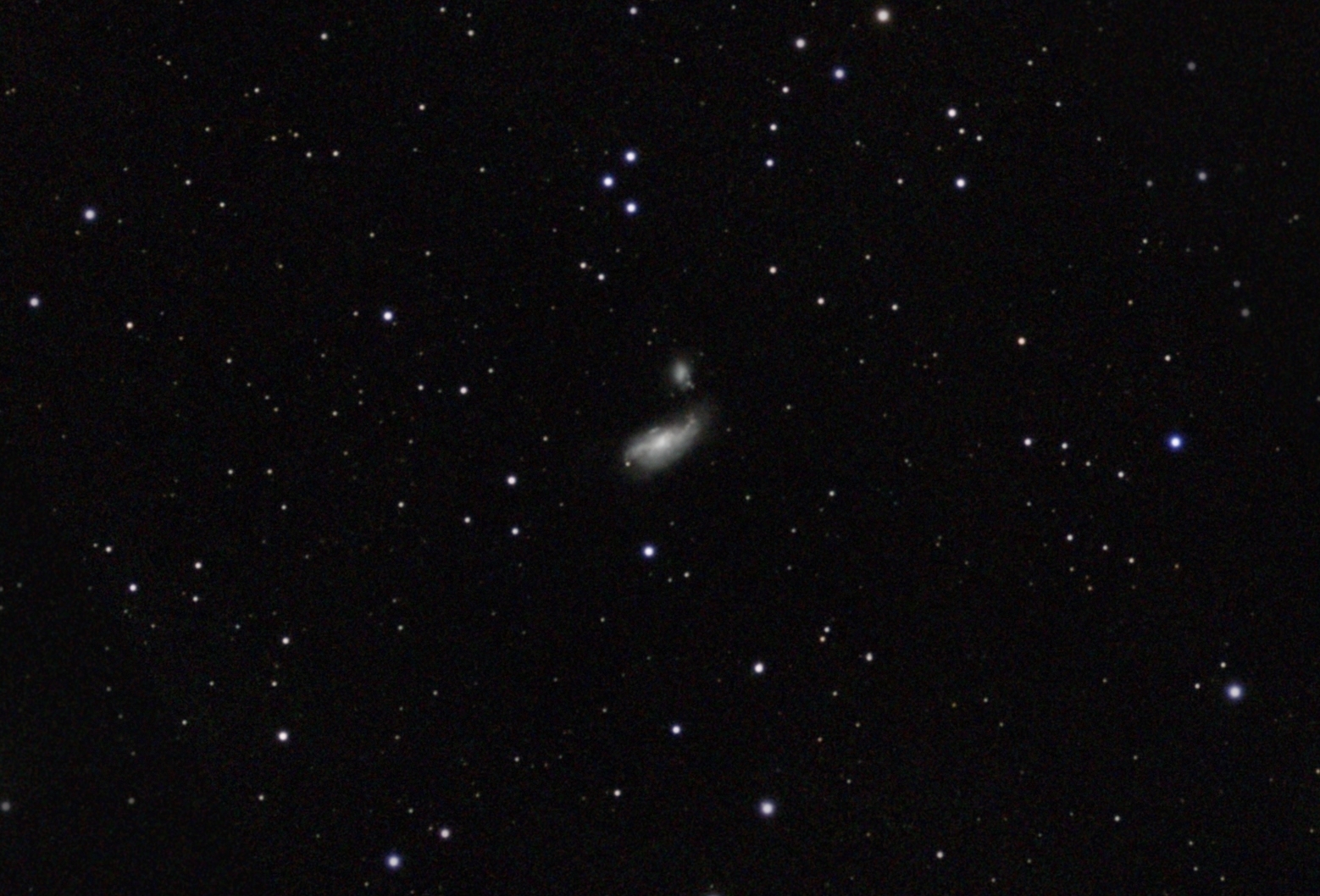}
    \caption{Composite image of the NGC4490 spiral galaxy, captured with a Stellina and a Vespera: 1590 frames of 10 seconds captured during several nights (9/6/2023, 23/6/2023, 18/3/2024, 3/4/2025).}
    \label{fig:ngc4490}
\end{figure}

\begin{figure}[htbp]
    \centering
    \includegraphics[width=0.95\linewidth]{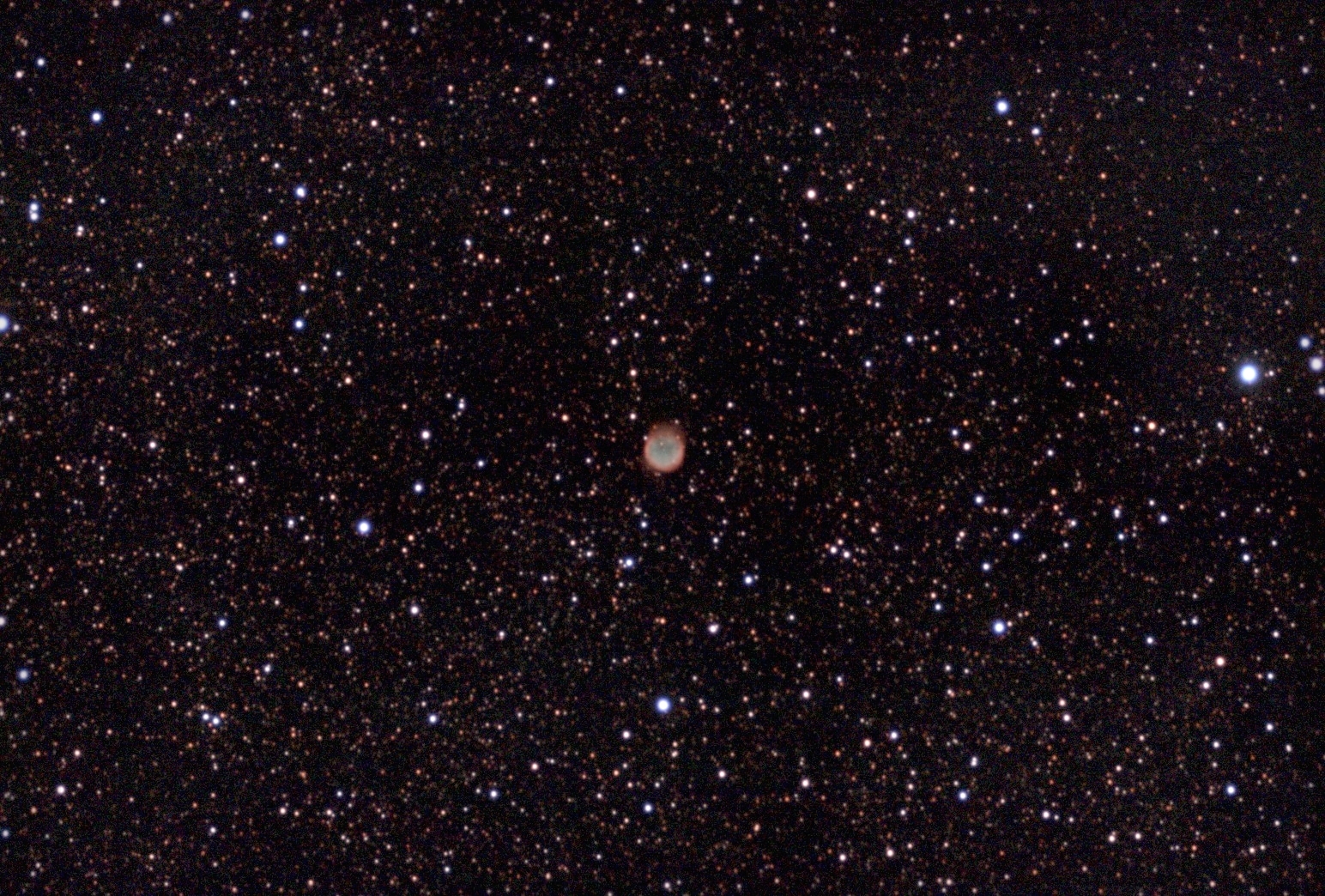}
    \caption{Composite image of the NGC6781 planetary nebula, captured with a Stellina and a Vespera: 1590 frames of 10 seconds captured during several nights (13/8/2022, 25/8/2023, 30/8/2023, 6/10/2023, 1/7/2024, 25/8/2025).}
    \label{fig:ngc6781}
\end{figure}

We have obtained hundreds of stacked images, including several images for targets that were frequently observed (e.g., 11 stacked images for M1).
Then, we have combined the images using a Python script relying on OpenCV, Pillow, and the Astroalign package \cite{beroiz2020astroalign}. 
The process have consisted of adjusting the image resolution, aligning them using the stars, and computing a weighted average of the images (with weights corresponding to the acquisition time of each individual image), followed by a slight stretch when possible. 
With targets having a lot of aligned images, this procedure notably allowed to obtain less noisy and more detailed images.
This was notably observed for Messier 1 (Figure \ref{fig:m1}), Messier 4 (Figure \ref{fig:m4}), NGC4490 (Figure \ref{fig:ngc4490}), NGC6781 (Figure \ref{fig:ngc6781}).

\subsection{Annotation}

The annotations on the images were made by drawing bounding boxes using the MakeSense software \footnote{\url{https://www.makesense.ai/}}, ensuring that the boxes accurately delineated the portions of the DSO that were effectively visible in the images.
The images were annotated at their native resolution to avoid losing detail. 

Bounding boxes for galaxies and planetary nebulae were relatively easy to define, as these objects are generally high-contrast. 
For star clusters (especially globular clusters), the annotations mainly focused on the dense core of the objects. 
For emission and reflection nebulae, we primarily delineated the most prominent regions while avoiding areas that could correspond to noise or background sky.
It dataset could naturally be designed as a multi-class dataset, given the diversity of the targets. 
However, we chose to focus primarily on the detection of objects of interest to astrophotographers

\section{Technical Validation}

To assess the impact of the proposed \textit{test2026} split, we have realized a preliminary evaluation the detection models presented in \cite{parisot2025robustnessanalysisdeepsky} on both the original test split and \textit{test2026}.
To this end, we have used the Ultralytics Python package \footnote{\url{https://www.ultralytics.com/}} with the standard hyper-parameters for each model archictecture, and we have ran the detection by resizing the high-resolution images of \textit{test2026}.

\begin{table}[h]
\centering
\caption{Preliminary performance comparison of trained models introduced in \cite{parisot2025robustnessanalysisdeepsky} on the different evaluation splits of DeepSpaceYoloDataset.}
\label{tab:results}
\begin{tabular}{|l|l|l|l|l|l|l|}
\hline
model   & split    & images count & precision   & recall   & mAP50   & mAP50-95 \\
\hline
yolo5m  & test     & 222          & 0.777       & 0.640    & 0.733   & 0.542      \\
yolo5m  & test2026 & 335          & 0.836       & 0.658    & 0.698   & 0.542      \\
\hline
yolo8m  & test     & 222          & 0.767       & 0.661    & 0.739   & 0.555      \\
yolo8m  & test2026 & 335          & 0.836       & 0.671    & 0.706   & 0.566      \\
\hline
yolo11m & test     & 222          & 0.802       & 0.619    & 0.738   & 0.554      \\
yolo11m & test2026 & 335          & 0.825       & 0.691    & 0.714   & 0.615      \\
\hline
yolo12m & test     & 222          & 0.776       & 0.727    & 0.784   & 0.604      \\
yolo12m & test2026 & 335          & 0.828       & 0.666    & 0.693   & 0.523      \\     
\hline
\end{tabular}
\end{table}

According to Table \ref{tab:results}, the results show that while some models perform very well on the original test set, their relative ranking changes when evaluated on the newer split. 
In particular, the model achieving the best results on the original test data does not maintain the same advantage on \textit{test2026}, whereas other models appear to generalize better to this newer dataset.
Overall, performance differences between the two splits suggest that \textit{test2026} introduces slightly more challenging or different data conditions. 
The models tend to produce more confident detections but show a small decrease in overall detection quality, which may indicate increased difficulty in accurately localizing or detecting all objects.
These observations suggest that the \textit{test2026} split likely reflects a shift in data distribution and provides a more demanding benchmark for evaluating model robustness and generalization to newer data.

Future work aimed at developing reliable detection models may use either split, with a preference for the new one, as it presents a greater diversity of data that are not present in the training and validation splits.

\section{Discussion}

DeepSpaceYoloDataset contains processed astronomical images that can be directly used for data analysis and machine learning applications. 
It is particularly well suited for the development and evaluation of Deep Learning approaches, including both supervised and unsupervised methods, as well as other computer vision techniques applied to astronomical imagery. Such data can support a variety of tasks, including object detection, representation learning, and automated analysis of deep-sky observations. 
In addition, the images may also be used for scientific purposes such as astrometric or photometric studies, complementing observations obtained through professional ground-based sky surveys.

\section{Conclusion}

In this work, we have introduced an updated version of DeepSpaceYoloDataset with the addition of the \textit{test2026} split, aimed at providing a more diverse benchmark for evaluating DSO detection models. 
By extending the dataset with new observations acquired under varied conditions, this update contributes to improving the robustness and evaluation of computer vision approaches applied to Electronically Assisted Astronomy and smart-telescope imagery. 
Although the dataset already covers a wide range of DSO, it is not intended to be exhaustive. 
In particular, it currently lacks objects from the southern celestial hemisphere. 

As future work, we plan to extend the dataset with observations of southern-sky targets and to explore a multi-class version of the annotations, enabling models to distinguish between different categories of DSO such as galaxies, nebulae, and star clusters. \\

\noindent \textbf{Acknowledgements}: This research was funded by the Luxembourg Institute of Science and Technology (LIST), during the NEOD2 research project (\url{https://researchportal.list.lu/projects/detail/neod}). The simulations were performed on the Luxembourg national supercomputer MeluXina. The authors gratefully acknowledge the LuxProvide teams for their expert support. \\ \\

\noindent \textbf{Data availability}: the different versions of DeepSpaceYoloDataset are available on Zenodo \url{https://doi.org/10.5281/zenodo.8387070}. Additional materials used to support the results of this paper are available from the corresponding author upon request.

\bibliographystyle{splncs04}
\bibliography{refs}

\end{document}